\documentclass[prb,twocolumn,showpacs]{revtex4}

\usepackage{graphicx}
\usepackage{amssymb}
\usepackage{epsf,latexsym}
\usepackage{times}

\newcommand{\ket}[1]{\left| #1 \right\rangle}

\newcommand{\be}{\begin{equation}}
\newcommand{\ee}{\end{equation}}
\newcommand{\ba}{\begin{eqnarray}}
\newcommand{\ea}{\end{eqnarray}}

\newcommand{\Fig}[1]{Fig.~\ref{#1}}
\newcommand{\Eq}[1]{Eq.~(\ref{#1})}

\begin{document}


\title{Entanglement distribution for a practical 
quantum-dot-based quantum processor architecture}

\author{Timothy P. Spiller$^{1}$}
\email{tim.spiller@hp.com}
\author{Irene D'Amico$^{2}$}
\email{ida500@york.ac.uk}
\author{Brendon W. Lovett$^{3}$}
\email{brendon.lovett@materials.oxford.ac.uk}

\affiliation{$^1$ Hewlett-Packard Laboratories, Filton Road, 
Stoke Gifford, Bristol BS34 8QZ, United Kingdom\\
$^2$ Department of Physics, University of York, York YO10 5DD, United Kingdom\\
$^3$ Department of Materials, University of Oxford, OX1 3PH, United Kingdom}

\date{\today }


\date{\today }

\begin{abstract}
We propose a quantum dot architecture for enabling universal quantum information processing. Quantum
registers, consisting of arrays of vertically stacked self-assembled 
semiconductor quantum dots, are connected by chains of in-plane self-assembled
dots. We propose an entanglement distributor, a device for producing 
and distributing maximally entangled qubits 
on demand, communicated through in-plane dot chains. This enables the
transmission of entanglement to spatially separated
register stacks, providing a resource for the realisation of a sizeable quantum processor
built from coupled register stacks of practical size.
Our entanglement distributor
could be integrated into many of the present proposals 
for self-assembled quantum dot-based quantum computation. 
Our device exploits the properties of simple,
relatively short, spin-chains and does not require microcavities.
Utilizing the properties of self-assembled quantum dots, after distribution 
the entanglement can be mapped into relatively long lived spin qubits and purified,
providing a flexible, distributed, off-line resource.
\end{abstract}

\pacs{03.67.Lx,78.67.Hc,85.35.-p,03.67.-a}


\maketitle 

\section{Introduction}
It is now well known that information technology (IT) that processes 
information according to quantum rules could have security and computational
advantages over conventional IT. Over the last decade there have been a great many
proposals for the building blocks of such new technology---how to realise qubits
and do gates between them---and the emergence of promising experimental results for
many of these proposals. The next steps towards useful QIT are further experimental
investigations, and theoretical considerations of outstanding practical issues. 
The work we present here is of the latter variety, showing how to distribute entanglement 
and thus connect together 
small quantum dot (QD) registers, of realistic size, to enable a larger, useful, 
quantum processor architecture.

In the past few years, self-assembled semiconductor QDs~\cite{QD} 
have been considered as one of the most promising solid state hardware routes 
for implementing quantum computation (QC)~\cite{PRLPRBBiolatti,PRBSergio,Briggs1,Briggs2,Mang1,Mang2,Paulibl}. 
Self-assembled QDs are characterized by different properties, dependent upon 
whether they are {\it stacked} or {\it in-plane}.
In the stacked arrangement, the QDs tend to be noticeably different in size
due to the strain propagation,
which makes them good candidates for energy-selective addressing of
excitations in  single QDs~\cite{PRLPRBBiolatti,PRBSergio,Paulibl}.
In-plane QDs can already be produced with fairly uniform size 
(with just a few percent size variation), driven by the crucial uniformity 
requirements for QD-based lasing~\cite{lasing}.  Furthermore, 1D QD chains with controllable QD size and density have been recently grown\cite{QDchain}   and {\it regular} 2D and 3D QD  arrays
can be now produced\cite{dotstructures}, creating real 
optimism for future QD technology.
For QC, the key feature of QD hardware is the possibility of optically driven 
quantum evolution, which could enable useful quantum processing inside the 
relatively fast decoherence times typical of solid state systems. 
In order to implement such QC schemes, arrays of vertically stacked 
quantum dots~\cite{PRLPRBBiolatti,PRBSergio,Briggs1,Briggs2,Mang2,Paulibl}
have been proposed as quantum registers. However, for practical reasons it is
clear that an individual 
stacked array cannot be scaled to an arbitrarily large qubit number. 
For actual technology, an architecture is needed in
which stacked registers of practical size are connected together, preferably in a 
way that is compatible with the manufacture of the registers themselves -- and this is
what we propose in this paper. We exploit recent results showing the feasibility of  quantum
communication between separated quantum registers, based on quantum buses 
made from chains of in-plain quantum dots and the 
properties of relatively short spin chains~\cite{IDA}. Our proposed architecture and how is used is described in the next section; we discuss the fidelity of its operation in Section~\ref{fidelity}, and conclude in Secton~\ref{conclusion}.

\section{The Entanglement Distributor}
\label{distributor}

\begin{figure}
 \vspace{1cm}
\hspace{2cm}\includegraphics[width=3.5in,height=2.0in]{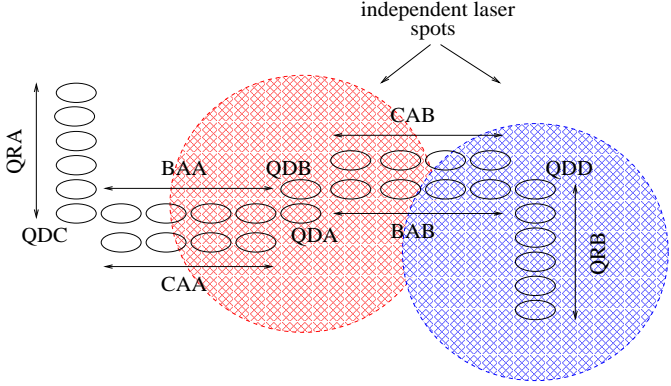}
\caption{Schematic entanglement distributor: Entangled excitons are produced in 
QDA and QDB, which are connected to in-plane dot quantum buses BAA and BAB respectively.
The transfer is controlled by dot control arrays CAA and CAB, as indicated. The buses connect
to dots QDC and QDD, which are located at the ends of the stacked registers QRA and QRB. These registers
and the dot-pair (QDA, QDB) can be addressed by different laser spots, as shown.}
\label{fig1}
\end{figure}

The key to our proposal is an entanglement distributor, an all-QD device for 
generating maximally entangled qubits on demand and distributing 
them to spatially separate regions. 
The system we envisage is shown schematically in in \Fig{fig1}.
Registers of stacked QDs are connected by entanglement distributors consisting of in-plane chains of QDs.

The entanglement distribution starts with the generation of a Bell state 
$|\psi\rangle = 2^{-1/2}(|00\rangle + |11\rangle)$ in QDA and QDB. 
For excitonic encoding, the computational basis is the 
absence ($|0\rangle_i \equiv |vac\rangle_i$) 
and presence ($|1\rangle_i \equiv |X\rangle_i$) of a ground state exciton in QD$i$.
Due to strain propagation~\cite{QD} the QD energy spectra differ in the vertical (stacked)
direction, i.e. between QDA and QDB, between each dot in the buses 
(BAA and BAB) and its partner dot in the control array (CAA and CAB), 
and along each of the quantum registers QRA and QRB (as shown in \Fig{fig1}). 
These energetically distinct dots may be addressed by laser pulses of different frequency.
Furthermore, since the excitons in vertically adjacent QDs have different energies,
resonant F\"orster coupling~\cite{Foerster1,Foerster2} is strongly inhibited in this direction.
The initial entangled state in QDA and QDB is generated by a two-colour 
laser pulse sequence~\cite{PRLPRBBiolatti}; this can be done on a picosecond time scale.
The first pulse is a $\pi/2$-pulse between the vacuum and ground state exciton in QDA. This is equivalent to a Hadamard operation $H_A$ applied to an empty QDA, 
$H_i |0\rangle_i = 2^{-1/2}(|0\rangle_i + |1\rangle_i)$.
(Note, for later use, that $H_i |1\rangle_i = 2^{-1/2}(|0\rangle_i - |1\rangle_i)$.) 
The second laser pulse performs a C-NOT gate on QDB, flipping QDB conditional  
on the presence of an exciton in QDA, due to the direct Coulomb coupling~\cite{PRLPRBBiolatti} between these excitons. Clearly the roles of QDA and QDB can be 
interchanged.

As mentioned above, in-plane QDs can be almost monodisperse,
so that we can assume adjacent QDs in each bus (BAA and BAB) and each 
control array (CAA and CAB) are very similar. This means that 
resonant F\"{o}rster energy transfer~\cite{Foerster1,Foerster2} takes place along the buses shown in \Fig{fig1}.
In addition, the relative stacked position of bus and control arrays
allows us to address the 
control arrays CAA and CAB (i) separately and (ii) without exciting
excitons also in the corresponding bus arrays.
Finally, the length of the bus arrays is made greater than the laser wavelength(s), 
so the central coupled  QD pair (QDA and QDB) and the two quantum registers
QRA and QRB can each be addressed with independent laser spots, 
as illustrated in \Fig{fig1}.

\begin{figure}
\vspace{1cm}
\hspace{3cm}\includegraphics[width=3.5in,height=2.8in]{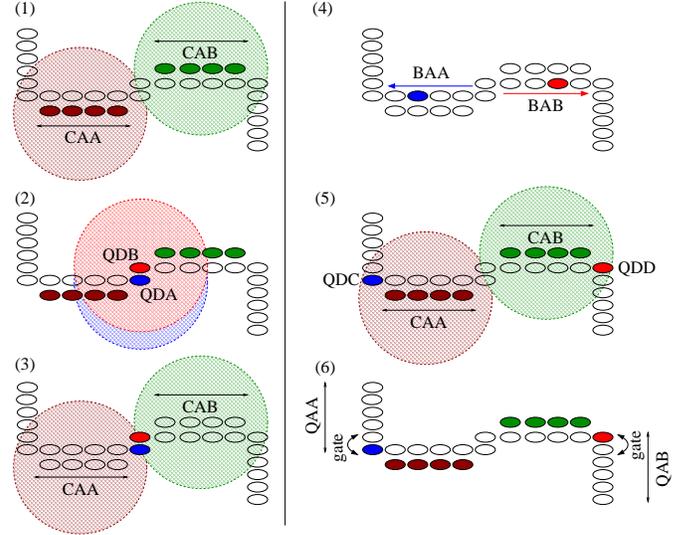}
\caption{The entanglement distributor: 
(1) The bus arrays BAA and BAB are set off resonance through the creation of an
exciton in each control array (CAA and CAB) QD. 
(2) The entangled pair is created in QDA and QDB. 
(3) The bus arrays BAA and BAB are set on resonance through the annihilation of the
excitons in control arrays CAA and CAB.
(4) The entangled qubits propagate down the buses.
(5) When the qubits reach QDs QDC and QDD, 
the bus arrays BAA and BAB are again set off resonance through the creation of an
exciton in each control array (CAA and CAB) QD.
(6) The entanglement is delivered to the registers QRA and QRB.}
\label{scheme}
\end{figure}

Once generated, the entanglement needs to be distributed.
The Hamiltonian for a chain of $N$ QDs, with zero or one exciton in each dot, is given by
\begin{eqnarray} H &=& \sum_{i=1}^N E_i |1\rangle_i\langle 1|_i \nonumber \\ &+&
 \sum_{i=1}^{N-1} \left( V_{{\rm F}i,i+1}|1\rangle_i\langle 0|_i\otimes |0\rangle_{i+1}\langle 1|_{i+1}+h.c.\right),\label{HF}\end{eqnarray}
where $V_{Fi,i+1}$ is the F\"orster coupling~\cite{IDA}.
This Hamiltonian governs the transfer of the entangled pair from QDA(B) to 
register QRA(B). For uniform in-plane chains, we shall assume that $E_i$ is independent of $i$ and $V_{{\rm F}i, i+1} = V_{\rm F}$.
The complete process of generating and distributing a maximally 
entangled qubit pair is shown in \Fig{scheme}, panels (1) to (6). 

\begin{figure}
\centering
\vspace{1cm}
\includegraphics[width=3in,height=4.2in]{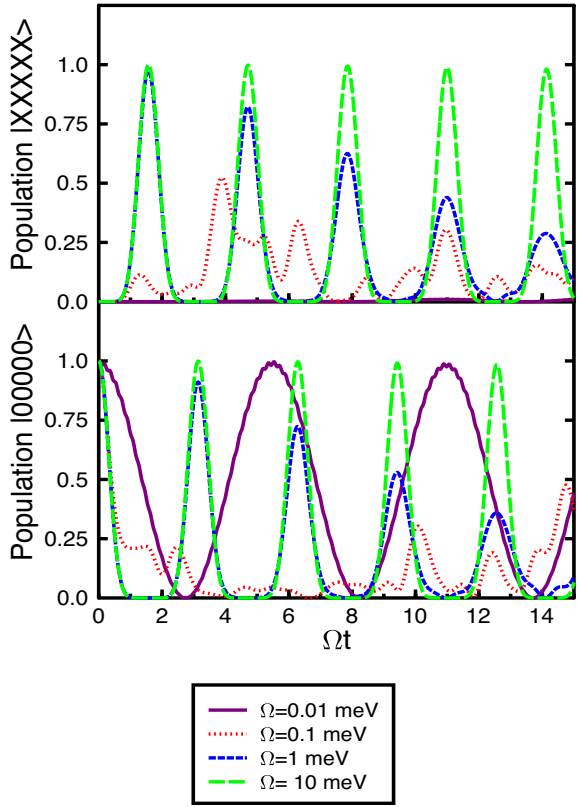}
\caption{Population of the ground state ($\ket{00000}$) and state with an exciton on every QD ($\ket{{\rm XXXXX}}$) in a 5 QD chain, for various values of dot-laser coupling $\Omega$. $V_{\rm F} = $0.2~meV for all plots.}
\label{control}
\end{figure}

In the first step (panel (1)) the bus arrays BAA and BAB are set off 
resonance with respect to excitations in QDA and QDB, by 
generating an exciton in each QD in the control arrays CAA and CAB. This multiple-excitation process, done with a single laser pulse, is not straightforward, and care must be taken when considering the system parameters.  
In our case $E_i \gg V_{\rm F}$ in Eq.~\ref{HF} and  the Hamiltonian has a series of eigenstate manifolds, each of which corresponds to a different total exciton number for the chain. The typical energy range covered by each manifold is on the order of $V_{\rm F}$.
Excitons are created in the dots by applying pulses centred on  energies 
$\hbar\omega_{cA}$ and  $\hbar\omega_{cB}$ that correspond to the $E_i$ for each chain. The coupling between laser and each individual dot is $ \Omega$ (which determines the Rabi frequency of the exciton creation process). For creation of an exciton on each dot $\Omega \gg V_{\rm F}$; in this case $\Omega$ becomes the dominant energy in the system and the dots behave as though they are uncoupled (to first order). We show how the population of the ground state of a 5-QD chain, and the population of the state with an exciton on all dots of a 5-QD chain varies as a function of time, and Rabi frequency, in Fig.~\ref{control}. Notice that, for realistic F\"orster couplings ($V_{\rm F} = $0.2~meV~\cite{Foerster2}), it is possible to drive the desired control-array dynamics $\ket{00000}\leftrightarrow\ket{{\rm XXXXX}}$ with very high accuracy on a sub-picosecond time scale. 

\begin{figure}
\vspace{1cm}
\hspace{3cm}\includegraphics[width=3in,height=2in]{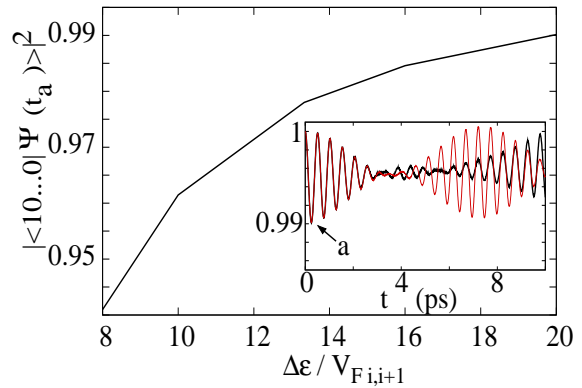}
\caption{Overlap between the evolution of the state (QDA + bus BAA + QDC)
 and its initial state, corresponding to (one exciton in QDA + empty bus BAA + empty QDC),
calculated at time $t=t_{\rm a}$, at which such an overlap is minimum.
 This is plotted as a function of the ratio
(biexcitonic shift/F\"{o}rster coupling), the dimensionless shift 
in the excitonic transition of a QD in
the bus BAA due to the presence of an exciton in the 
corresponding QD in control array CAA. Inset: The time evolution of the overlap
for a ratio of 20, for the cases of five (thin line) and seven (thick line) QDs in (QDA+BAA+QDC).
The difference in chain length only has an effect after several
oscillations, and this is typical for a dimensionless shift $\gg 1$. Point `a' marks
the minimum used in the main figure for calculating the overlap.}
\label{Forstblock}
\end{figure}

The excitons created in the control bus couple to the excitonic transitions in BAA and BAB, 
causing a biexcitonic energy shift $\Delta E_{XX}$ and effectively detuning 
them from resonance with respect to QDA and QDB. 
This shift can be tailored up to several meV~\cite{PRLPRBBiolatti}, 
inhibiting F\"orster processes between QDA (QDB) and 
the QDs forming the bus BAA (BAB). This inhibited dynamics is shown in 
\Fig{Forstblock}. When bus BAA is off resonance, due to the excitons in control array 
CAA, an initial state of an exciton in QDA (with bus BAA empty) remains
in that state with at least 99\% 
probability as it evolves, for a ratio $\Delta E_{XX}/V_{\rm F}= 20$. This ratio could be achieved with 
a shift of 4 meV and a F\"{o}rster coupling of 0.2 meV, so very good confinement
of an exciton is possible for practical parameters.

The second step (panel 2) is to generate a maximally entangled pair of excitons, 
one in QDA and one in QDB.
Since QDA and QDB are stacked QDs, they can be individually addressed using sub-ps laser pulses of different color, and creation of the required state proceeds through Hadamard and C-NOT gates as we discussed earlier. 
Owing to the confinement discussed above, the entangled excitons do not propagate along the buses.

In the third step (panel 3), the bus arrays BAA and BAB are again set into 
resonance with the excitations in QDA and QDB. This is done by recombining the 
excitons in each QD in the control arrays by the same laser pulses as 
those used in step 1.
Now the entangled excitons in QDA and QDB are free to 
travel along BAA and BAB, respectively (panel 4); the time evolution follows the dynamics of 
relatively short chains described in Ref.~\cite{IDA} and governed by \Eq{HF}.
In plane QDs have nearest-neighbor couplings that are typically {\it un-modulated}.
Chains of such QDs can be used for state transfer, and this avoids complexity and adds to the practicality of our scheme;
the dynamics of exciton transfer down these chains has 
been studied in Ref~{IDA}. These dynamics are characterized by a series of resonances,
corresponding to different fidelities of the transfer, the very first of which has 
a fidelity greater than 94\% for the chains required for our
entanglement distributor (which is less that 10 QDs long). 
The actual transfer time depends on the strength of the F\"orster 
coupling between in-plane QDs, and it is foreseen that  such QD 
structures can be optimized to have a F\"orster coupling of the order 
of a meV~\cite{Foerster2} -- and hence a time of transfer between QDA(B) and QDC(D) 
of the order of a picosecond~\cite{IDA}. For the example discussed earlier (where  
$V_{\rm F} =$ 0.2 meV) the transfer time is still less than 10~ps for a bus with 9 QDs.
We underline that this transfer time is, on the one hand, much longer than the characteristic time corresponding to the control-array dynamics (whose population/de-population can therefore be viewed as instantaneous), but on the other hand it is still much shorter than the relevant decoherence times (of the order of a nanosecond), allowing for the possibility of further manipulations. 

When the time elapsed is equal to the transfer time~\cite{IDA}, 
the two entangled excitons will be in QDC and QDD, i.e. at the bases 
of the two different quantum registers QRA and QRB.
Using laser pulses to once again set the buses  BAA and BAB out of resonance 
with respect to QDA, B, C and D, will confine the two entangled 
excitons in QDC and QDD, (panel 5). 
Thus the entangled pair has now been distributed to spatially separated 
regions, which can be addressed separately, e.g. by using laser pulses.
Using optical control, quantum operations can be performed on such a pair (panel 6) 
and the entanglement can be transferred to different degrees of freedom,
e.g. other excitons, spatially distinguishable photons, 
or electronic spins. These may be characterized by much longer decoherence times than the original excitons.

As an example of this transfer, we outline how a qubit encoded in the
exciton basis in a QD can be swapped into the spin of an excess electron 
in an adjacent Pauli-blockade QD. In terms of the spin, the qubit
computational basis is $|0\rangle \equiv |\uparrow\rangle$ and
$|1\rangle \equiv |\downarrow\rangle$. In a Pauli-blockade QD ~\cite{Paulibl}
a suitably polarized optical $\pi-$pulse will create 
(or remove) an exciton {\it conditional} 
on the spin state of the excess electron, so, for example, a qubit with its ancillary exciton state is 
represented as $|0\rangle \equiv |\uparrow,vac\rangle$ and
$|1\rangle \equiv |\downarrow,X\rangle$ in such a QD after conditional creation.
If such a QD is adjacent to another QD (QDC or QDD in our scheme) containing an
excitonic qubit, a two-qubit gate equivalent to a controlled-phase gate (which
flips the sign of the $|11\rangle$ computational basis amplitude and leaves the others alone) can
be achieved ~\cite{Paulibl} using the biexcitonic shift, through conditional creation of an
exciton for a chosen time. We denote this gate by $P_{ij}$ where $i$ labels the dot 
C or D and $j$ labels the appropriate adjacent Pauli-blockade QD containing the 
spin qubit.
If this latter qubit is set to a known initial state, say $|0\rangle$, a SWAP 
(denoted by $S_{ij}$) between
the two qubits can be achieved through two C-NOT gates, which decomposes to 
$ S_{ij} = H_i P_{ij} H_i H_j P_{ij} H_j \; .$
Through this gate sequence an excitonic qubit that has been distributed to QDC or QDD
can be swapped into a (relatively) long-lived spin qubit in an adjacent QD. We note that the 
two Hadamard operations $H_j$ have to be performed on the spin qubit. However,
fast optical techniques exist for performing such gates ~\cite{Paulibl, brendon05}, so in principle
the whole SWAP operation can be done on a fast optical/excitonic timescale (i.e. no more than a few ps), to
mediate against the effects of decoherence acting on the excitons.

\section{Fidelity of the Distribution Process}
\label{fidelity}

From a practical quantum processing perspective, the swapping of the entangled qubits
into spins, to create a long-lived entangled resource, is very important. The fidelity of our distribution scheme is set by a product of factors. First, the chain transfer fidelity of 94\%, second  
the control chain blocking fidelity of around 99\%, third, the SWAP gate fidelity -- which can also be greater than 99~\%~~\cite{brendon05} -- and fourth, the natural decoherence of excitons due to spontaneous emission of photons, or exciton-phonon interactions~\cite{otherfactors}. The decoherence of excitons in QDs consists of two main features: an initial sub-picosecond pure dephasing (caused by the excitons' interaction with acoustic phonons), followed by a decay of the remnant polarization that decays through exciton recombination. (See Refs.~\cite{bayer02},~\cite{Kuhn02},~\cite{Kuhn} and \cite{borri04} and for a comprehensive description of these effects and a comparison of experimental data with theory).


However, for the purposes of our discussion, we will take the exciton recombination to be the principal source of dephasing. This is for two main reasons. First, The value of the residual polarization  strongly depends on temperature and material parameters but is generally much larger than the lost polarization. Second, the initial decay corresponds to the formation of exciton-phonon dressed states, and we could redefine our qubit to correspond to one of these. This would effectively eliminate the fast initial decay.




We estimate that the total time for one entanglement distribution operation is no more than 20~ps, whereas the exciton decay time is about 1~ns~\cite{borri01} (giving a `natural' fidelity of around 99\%). Thus, an estimate of the overall fidelity of distribution is 
91\%. Such a good fidelity is certainly adequate for initial experimental investigations. 
It could be used to demonstrate teleportation
and it might even be useful in certain few-qubit applications, which are run
probabilistically with rather limited qubit resources. However, for practical
quantum processing and a scalable processing architecture, very high fidelity 
entanglement is desirable. This could be achieved by purification~\cite{purify}.
A number of good fidelity pairs can be distilled to a smaller number of higher fidelity
pairs using local qubit operations. This is feasible if distributed excitonic
entanglement is mapped into long-lived spin entanglement prior to purification, as we propose.
Building up the entanglement resource {\it off-line} and using it to link together 
separate parts of the quantum device clearly has a practical advantage 
over trying to propagate a computation directly from one register to another, 
since it prevents errors caused by imperfect 
gate operations.

\section{Summary and Conclusion}
\label{conclusion}

We have presented a practical all-quantum-dot architecture for quantum
information processing. The key ingredient to this is an entanglement
distributor, which provides a resource of distributed entangled states on demand. This
enables the connection of stacked QD processors, of modest and practical
size, to form a larger quantum processor. Entangled exciton states are
transported to different regions of a semiconductor based quantum 
information device using chains of in-plane QDs, at which point they can, for example,
be swapped into long-lived spin qubits, or used to create entangled photons. 
The fidelity of the distributed entanglement is good, and could be further increased through 
purification. Such entanglement is a generic and flexible resource. 
For example, it could be used for teleportation
of quantum states, to enable distributed quantum gates, 
or to build distributed cluster states~\cite{cluster}, so it can be utilised in a variety
of approaches to quantum processing.
Our QD-based entanglement distributor should integrate with and thus enhance 
many of the current QD-based quantum information technology proposals. Given the
impressive progress with QD structures~\cite{lasing,dotstructures}, 
entanglement distribution forms a
realistic experimental goal in the relatively short term and a route to scalable solid state
QC in the long term.

BWL acknowledges support from the QIPIRC www.qipirc.org (GR/S82176/01), DSTL, St Anne's College, Oxford and the Royal Society.



\begin{thebibliography}{99}
\bibitem{otherfactors} We neglect three further decoherence mechanisms as they act only on a much slower timescale: single particle and exciton tunneling (see~\cite{lovett05})  and nuclear spin interactions (see~\cite{petta05}).
\end{thebibliography}
\end{document}